\documentclass[runningheads]{llncs}
\usepackage{graphicx}
\usepackage{verbatim}
\usepackage{rotating}
\usepackage{caption}
\usepackage{subcaption}
\usepackage{amssymb}
\begin{document}
%

\title{Benchmarking Hashing Algorithms for Load Balancing in a Distributed Database Environment\thanks{This paper was accepted to the 11th International Conference on Model and Data Engineering (MEDI'22).}}
\titlerunning{Benchmarking Hashing Algorithms for Load Balancing}

%
\author{Alexander Slesarev\inst{1,2}\orcidID{0000-0002-7109-3035} \and
Mikhail	Mikhailov\inst{1}\orcidID{0000-0002-8059-6769} \and
George Chernishev\inst{1,2}\orcidID{0000-0002-4265-9642}}

\authorrunning{A. Slesarev et al.}

%
\institute{Unidata \and Saint-Petersburg State University
\\
\email{alexandr.slesarev@unidata-platform.org, mikhail.mikhailov@unidata-platform.ru, 	georgii.chernyshev@unidata-platform.ru}}

\maketitle              
\begin{abstract}

Modern high load applications store data using multiple da\-ta\-ba\-se instances. Such an architecture requires data consistency, and it is important to ensure even distribution of data among nodes. Load balancing is used to achieve these goals.

Hashing is the backbone of virtually all load balancing systems. Since the introduction of classic Consistent Hashing, many algorithms have been devised for this purpose.

One of the purposes of the load balancer is to ensure storage cluster scalability. It is crucial for the performance of the whole system to transfer as few data records as possible during node addition or removal. The load balancer hashing algorithm has the greatest impact on this process.

In this paper we experimentally evaluate several hashing algorithms used for load balancing, conducting both simulated and real system experiments. To evaluate algorithm performance, we have developed a benchmark suite based on Unidata MDM~--- a scalable toolkit for various Master Data Management (MDM) applications. For assessment, we have employed three criteria~--- uniformity of the produced distribution, the number of moved records, and computation speed. Following the results of our experiments, we have created a table, in which each algorithm is given an assessment according to the abovementioned criteria.

\keywords{Consistent Hashing  \and Databases \and Benchmarking.}
\end{abstract}

\section{Introduction}

As any organization grows, the volume of its corporate data assets rises as well. There are two general approaches to solving this architectural problem~\cite{10.5555/1972515}: vertical and horizontal scaling. Vertical scaling focuses on increasing the capabilities of a single server, whereas horizontal scaling involves adding machines to the cluster. To implement horizontal scaling, a database table has to be horizontally split into parts (shards), which are stored on different server nodes.

Horizontal scaling has several significant advantages, such as the possibility to flexibly adjust storage volume by altering cluster size. Another one is the ability to deal with data loss by replicating data among servers. Thus, the need for distributed data storage appears.

An important component of distributed storage is the load balancer~--- a mechanism which determines which particular server will store a data entity (e.g. record, table part, etc). There are several assessment criteria for load balancers. First of all, data distribution over servers should be as close to uniform as possible. Next, if cluster size changes, the number of moved data entities must be close to optimal. And finally, load balancer computing costs should not be high.

In order to calculate the shard assigned to a given data entity, the load balancer utilizes a hashing algorithm. Since the 90's, many hashing algorithms have been designed specifically for balancing different types of loads such as network connection management, distributed computing optimization, and data storage balancing.

One of the research disciplines focusing on storage and processing of large data volumes is Master Data Management~\cite{AllenCervo,10.5555/1457711} (MDM). It is based around the concept of Master Data~--- a concept that combines objects important for business operations within an organization, such as inventory, customers, and employees. The main goals of MDM are unification, reconciliation, and ensuring completeness of corporate Master Data.

In this paper, we compare several hashing algorithms and assess their applicability to the load balancing problem. We experimentally evaluate them using both simulated and real tests. For the latter, we employ the Unidata platform~\cite{DataPlat}~--- an open-source MDM toolkit that has distributed storage capabilities. Following the results of our experiments, we have created a table, in which each algorithm is given an assessment according to the abovementioned criteria.

This paper is organized as follows. In Section~\ref{sec:relwork} we describe a number of existing load balancing algorithms, define several terms from the MDM area, and review Unidata storage architecture. Then in Section~\ref{sec:evaluation} we describe the conducted experiments, and discuss the achieved results in Section~\ref{sec:discussion}. We conclude this paper with Section~\ref{sec:concl}.

\section{Background and Related Work}\label{sec:relwork}

In this section we describe those existing hashing algorithms that we are going to benchmark, and since our last series of experiments is run on a real system, we also provide a general description of the system itself, its purpose, and the used data schema.

This study concerns two research fields that have a rich body of work~--- load balancing and hashing algorithms. The former has a large number of surveys describing dozens of works. For example, consider study~\cite{JAFARNEJADGHOMI201750}, which references many more similar surveys. Unfortunately, such surveys only classify the covered methods using some high-level criteria (e.g. adaptivity, static or dynamic, preemptiveness, etc). They do not experimentally evaluate surveyed algorithms. The reason for this is the fact that such surveys are too broad, their reviewed studies belong to many different fields and it will be extremely difficult to run such an evaluation. At the same time, industry is interested in the best method for a particular domain, and the answer can be found only empirically.

Turning to hashing, we must mention a very comprehensive survey~\cite{10.1145/3047307}, which describes many hashing approaches, and proposes an algorithm taxonomy. However, the section that concerns data-oriented hashing is aimed towards data structures and machine learning, but not load balancing. The set of hashing algorithms that we consider in our study is absent in this survey.

Therefore, our work fills the gap in existing studies.

\subsection{Considered Methods}

Let us start with the hashing algorithms which are used for data balancing. Each of the considered load balancers applies its hash function to some incoming data entity, so we call the result of the application a \textit{key}. The purpose of a load balancer is to match each key to one of the shards, which is represented by an integer number (id). There are several hashing methods that we consider in this paper:
\begin{itemize}
    \item \textbf{Linear Hashing} is one of the overall oldest algorithms. Besides the classic version~\cite{LH} there is a number of modifications such as LH*~\cite{LH*}, LH*M~\cite{LH*M}, LH*G~\cite{LH*G}, LH*S~\cite{LH*S}, LH*SA~\cite{LH*SA} and LH*RS~\cite{LH*RS}. The core idea of algorithms of this family is to calculate the remainder of dividing the key by a number of shards in the system. Therefore, they are suitable for solving the problem featuring a fixed number of shards, which in our case is a disadvantage. In our work we have adopted a version used for partitioning in PostgreSQL\footnote{https://github.com/postgres/postgres/blob/master/src/backend/partitioning/partbounds.c}. 
    
    \item \textbf{Consistent Hashing}. Originally, this algorithm~\cite{ConsistentHash} was designed for balancing loads of computer networks. Nowadays, it seems to be the most popular method for balancing many various types of loads. For example, distributed systems like AWS DynamoDB~\cite{DynamoDB} and Cassandra~\cite{Cassandra} use Consistent Hashing for partitioning and replication. This method is based on picking random points on a ring, which is a looped segment of real numbers that represents shards and data entities as points. Points denoting keys are assigned to the clockwise nearest shard. To ensure even data distribution, each shard is represented by several points. Note that the design of this method allows to change the number of shards while moving only the optimal number of records.
    
    \item \textbf{Rendezvous}. Similarly to Consistent Hashing, this method~\cite{Rendezvous} was also developed to optimize network load. For a given key, the algorithm calculates the value of the cost function for each of the shards and assigns the key to the shard with the highest value. When adding or removing shards, Rendezvous also does not move extra records.
    
    \item \textbf{RUSH}~\cite{RUSH} was developed to store data in a disk cluster. It has two modifications: RUSH\textsubscript{R} and RUSH\textsubscript{T}~\cite{RUSHT}. Authors of RUSH focused on improving the uniformity of data distribution when changing the size of the cluster, therefore, the algorithm is based on the following principle: every time cluster size is changed, a special function is used to decide which objects should be moved to balance the system.
    
    \item \textbf{Maglev} is an algorithm~\cite{Maglev} from Google's load balancer for web services. The goal of Maglev is to improve data uniformity (compared to Consistent Hashing) and cause ``minimal disruption'', e.g. if the set of shards changes, data records will likely be sent to the same shard where they were in before. It was proposed as a new type of Consistent Hashing, in which the ring is replaced with a lookup table by which a key can be assigned to a shard. The size of the lookup table should be greater than the possible number of shards to decrease collision rate. Average proposed lookup time is $O(Mlog(M))$, where $M$ is size of lookup table.

    \item \textbf{Jump} is another load balancer from Google~\cite{Jump}. Its authors presented it as a superior version of Consistent Hashing which ``requires no storage, is faster, and does a better job of evenly dividing the key space among the buckets''. Jump generates values for shard numbers only in the $[0; \# shards]$ range, so the addition of a new shard is fast. However, a deletion of an intermediate shard will cause rehashing of many records. It takes $O(log(N))$ time to run Jump, where $N$ is the number of shards.
    
    \item \textbf{AnchorHash}. According to the authors, AnchorHash~\cite{AnchorHash} is a ``hashing technique that guarantees minimal disruption, balance, high lookup rate, low memory footprint, and fast update time after resource additions and removals''. A notable difference of AnchorHash from other discussed algorithms is that it stores some information about previous states of the system.
\end{itemize}

Each paper that proposed a novel hashing algorithm compared it only to a small number of other such algorithms. To the best of our knowledge, there were no dedicated comparisons of such algorithms applied to the horizontal scaling problem. At the same time, ensuring high performance of horizontal scaling is a pressing problem that is in demand by the industry. Thus, there is a need to evaluate all of these algorithms and study their applicability to this problem.

\subsection{Basic definitions}

To understand the specifics of the data storage in which the load balancers will be evaluated, it is necessary to introduce some MDM terms:

\begin{itemize}
    \item \textbf{Golden Record}. One of the main problems in MDM is the compilation and maintenance of a ``single version of the truth''~\cite{AllenCervo} for a given entity, e.g. person, company, order, etc. To achieve such a goal, an MDM system has to assemble information from many data sources (information systems of a particular organization) into one clean and consistent entity called the golden record.
    \item \textbf{Validity Period} is a time interval in which the information about an entity is valid. 
    For each golden record, several validity periods may exist. This fact should be taken into account while querying the data. There are two temporal dimensions: time of an event and time of introduction of this new version of information to the system. This leads to a special storage scheme for managing this information.
\end{itemize}

\subsection{System architecture}

MDM systems are a special class of information management systems~\cite{DataPlat}. Their specifics impose requirements on the platform storage architecture and data processing.

First of all, \underline{versioning} of stored objects must be supported. For this reason, data assets describing stored objects have validity periods and they should be taken into account while querying the data.

Second, deletion operations can be performed only by an administrator, while a user can only mark data entity as removed. This is necessary to avoid information loss and ensure correct versioning support. Sometimes there are legal requirements for this data handling semantics. Such an architectural pattern is often called \underline{tombstone} delete. 

Third, \underline{provenance} should be provided. It means that any system operation should be traceable. For example, there must be a way to roll back all changes in records after each operation. 

The proposed approach is based on the following four tables, where three of them represent entities:

\begin{itemize}
    \item Etalon stores the metadata of the golden record itself.
    \item Origin stores the metadata related to the source system of the record.
    \item Vistory (version history) is the validity period of origin, which in turn may have revisions.
    \item External Key is a table needed for accessing the data from within other parts of the Unidata storage.
\end{itemize}

\begin{figure}
    \centering
    \includegraphics[width=0.9\textwidth]{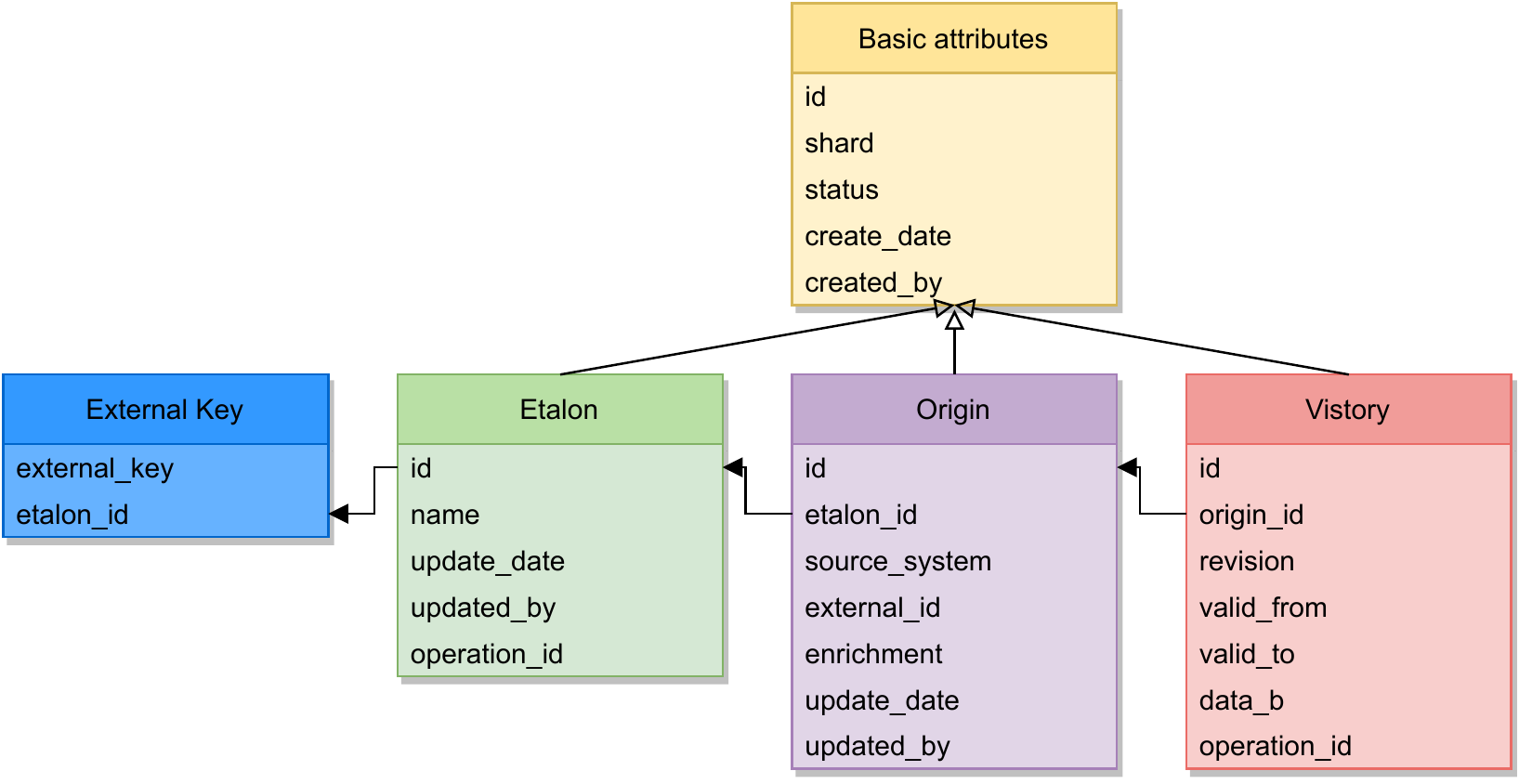}
    \caption{Tables used for data storage in the Unidata platform}
    \label{fig:tables}
\end{figure}

The relations between these tables are shown in Figure~\ref{fig:tables}. The links with empty arrowheads denote ``shared'' (inherited) attributes and full arrowheads show the PK-FK relationship. The detailed descriptions of table attributes can be found in~\cite{DataPlat}.  

\section{Evaluation}\label{sec:evaluation}

In order to select the best hashing algorithm for the load balancing problem, we have performed an experimental evaluation.

\subsection{Experimental Setup}

Experiments were performed using the following hardware and software configuration:
\begin{itemize}
    \item Hardware: LENOVO E15, 16GiB RAM, Intel(R) Core(TM) i7-10510U CPU @ 4.90GHz, TOSHIBA 238GiB KBG40ZNT.
    \item Software: Ubuntu 20.04.4 LTS, Postgres 11.x, JDK 11.x, Tomcat 7.x, Elasticsearch 7.6.x.
\end{itemize}

Some algorithms have parameters that affect their performance:
\begin{itemize}
    \item For Consistent Hashing, 16 points for each shard were selected. This number was chosen experimentally, as a compromise between the hashing speed and uniformity of the initial distribution.
    \item For Maglev, lookup table size was set to 103. Similarly to Consistent Hashing, this number was selected experimentally. Note that this value is important in rebalancing process, but not for lookup.
    \item For AnchorHash, we have set the $|\mathcal{A}|$ (the number of buckets that algorithm works with) to double the maximum number of shards (i.e 64) as it was recommended in the original paper~\cite{AnchorHash}.
\end{itemize}

\subsection{Results}

In order to evaluate the load balancing algorithms, we have defined three criteria which we ranked by their importance (in descending order):
\begin{enumerate}
    \item Uniformity of the produced data distribution.
    \item Redundant movement of records during shard addition or removal.
    \item Lookup speed.
\end{enumerate}

To select the best algorithm, we have conducted three experiments. First, we have performed a load balancing simulation experiment in Google Colab\footnote{https://colab.research.google.com/drive/1pbJUFFP9JsSTSn7nrWv0tYdiUg2uRxAv? usp=sharing} (in Python). This step was needed to run a shallow, preliminary assessment of algorithm performance. It was conducted as follows: first, 10K records were generated and distributed (via hashing function) into 32 shards. Thus, uniform distribution will result in 312 records per shard. After this, 8 shards were scheduled for removal and system was forced to rebalance the data. Therefore, uniform distribution should result in 416 records per shard. This procedure was run for each of the considered load balancers (hashing functions).

The mean values of ten such experiments are presented in Table~\ref{tbl:experiment1}. The first column of the table contains the average time of shard calculation, the second and the third present variance of records assigned to shards before and after rebalancing, respectively. The last column shows the ratio of the number of actually moved records to the optimal number.  

\begin{table}[]
\centering
\caption{First experiment, simulation in Colab}
\label{tbl:experiment1}
\begin{tabular}{l|l|l|l|l}
\multicolumn{1}{c|}{Algorithm} &
  \multicolumn{1}{c|}{\begin{tabular}[c]{@{}c@{}}Shard id calculation \\ time (ns)\end{tabular}} &
  \multicolumn{1}{c|}{\begin{tabular}[c]{@{}c@{}}Variance \\ before drop\end{tabular}} &
  \begin{tabular}[c]{@{}l@{}}Variance\\ after drop\end{tabular} &
  \begin{tabular}[c]{@{}c@{}}Moved records\\ ratio\end{tabular} \\
  \hline \hline
Consistent & 55049  & 72 & 94  & 1.00      \\ \hline

Rendezvous & 105331 & 18 & 21  & 1.00      \\ \hline
RUSH\textsubscript{R}      & 547044 & 95 & 125 & 1.57 \\ \hline
Maglev     & 1146   & 16 & 21  & 1.39 \\ \hline
Jump       & 18077  & 21 & 29  & 3.63 \\ \hline
AnchorHash~ & 3539   & 17 & 20  & 1.00    \\
\end{tabular}
\end{table}

Based on the experimental results, we have decided to exclude RUSH\textsubscript{R} from further consideration due to it failing to conform to all three criteria. We have also excluded Jump due to  the poor quality of rebalancing.

Our next experiments involve the Unidata platform, which is implemented in Java. To verify the consistency and transferability of previously obtained results, we have decided to re-evaluate the shard id calculation time inside the platform. Therefore, the second experiment was also to distribute 10K records among shards. The measured averages were as follows:
\begin{itemize}
    \item Linear Hashing~--- 808ns
    \item Consistent Hashing~--- 2419ns
    \item Rendezvous~--- 5945ns
    \item Maglev~--- 807ns
    \item AnchorHash~--- 2015ns
\end{itemize}

As one can see, the order of algorithm run times has not changed compared to the previous experiment. Therefore, we can conclude that switching the programming language did not affected the results of the previous experiment and we can continue using the Unidata platform.

\begin{figure}[!ht]
    \centering
    \includegraphics[width=\textwidth]{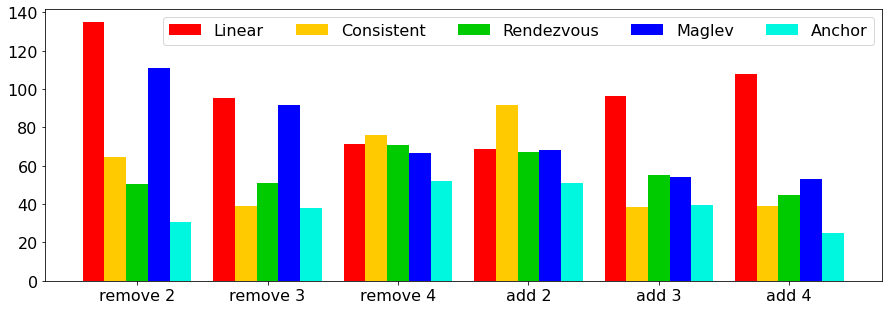}
    \caption{Rebalance time (seconds).}
    \label{fig:time}
\end{figure}

\begin{figure}[!ht]
\centering
    \centering
    \includegraphics[width=\textwidth]{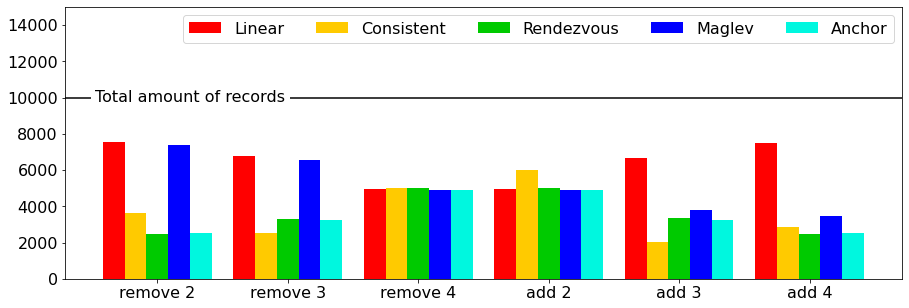}
    \caption{Number of records from the Etalon table that were moved in the rebalancing process.}
    \label{fig:moved-etalons}
\end{figure}

The third experiment was performed using a deployed Unidata platform. Its storage configuration was the following: four Docker nodes with PostgreSQL with eight shards on each node. We have generated 10K external keys and etalons as a workload. The idea of the experiment was as follows: remove three nodes one by one and then add them back in a similar manner.

\begin{figure*}
\centering
\begin{subfigure}{.99\textwidth}
    \centering
    \includegraphics[width=0.99\textwidth]{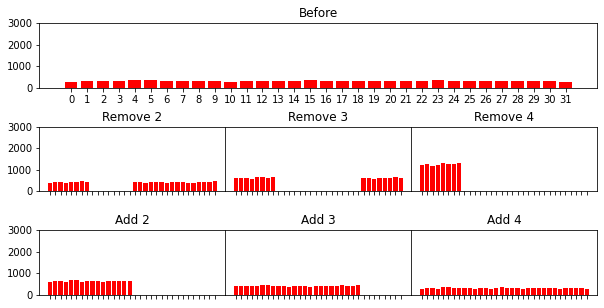}
    \caption{Linear Hashing.}
    \label{fig:linear}
\end{subfigure}
\begin{subfigure}{.99\textwidth}
    \centering
    \includegraphics[width=0.99\textwidth]{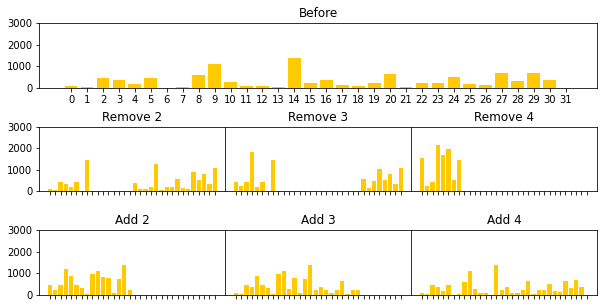}
    \caption{Consistent Hashing.}
    \label{fig:consistent}
\end{subfigure}
\begin{subfigure}{.99\textwidth}
    \centering
    \includegraphics[width=0.99\textwidth]{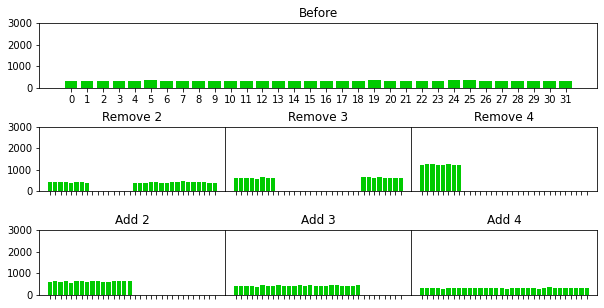}
    \caption{Rendezvous.}
    \label{fig:rendezvous}
\end{subfigure}
\caption{Distribution of Etalon table records among shards.}
\label{fig:distribution1}
\end{figure*}

\begin{figure*}
\centering
\begin{subfigure}{.99\textwidth}
    \centering
    \includegraphics[width=0.99\textwidth]{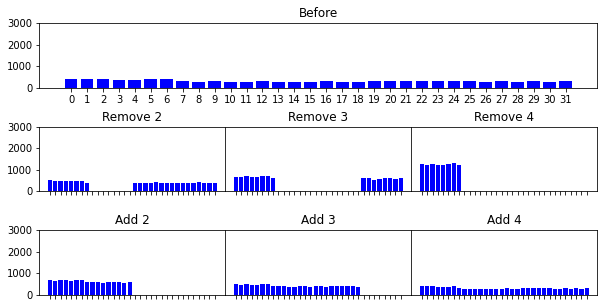}
    \caption{Maglev.}
    \label{fig:maglev}
\end{subfigure}
\begin{subfigure}{.99\textwidth}
    \centering
    \includegraphics[width=0.99\textwidth]{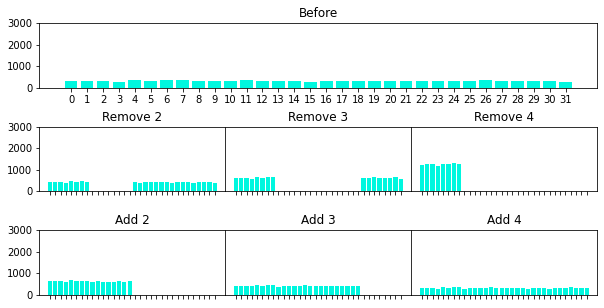}
    \caption{AnchorHash.}
    \label{fig:anchor}
\end{subfigure}
\caption{Distribution of Etalon table records among shards.}
\label{fig:distribution2}
\end{figure*}

The results of the evaluation are displayed in the following figures. Total time spent on each rebalance step is shown in Figure~\ref{fig:time}, and the number of moved etalons is shown in Figure~\ref{fig:moved-etalons}. We have omitted such figure for external keys since it is largely the same (it is 1:1 mapping). Data distribution among shards on each step is shown in Figures~\ref{fig:distribution1} and~\ref{fig:distribution2}.

This experiment allows us to draw the following conclusions:
\begin{itemize}
    \item Consistent Hashing, Rendezvous and AnchorHash move more than 50\% records less than Linear Hashing.
    \item During the first two rebalancing steps Maglev moved approximately the same number of records as Linear Hashing, but on the last two steps Maglev moved significantly less records and got close to the other three methods.
    \item Linear Hashing, Rendezvous, Maglev and AnchorHash distribute data uniformly enough, but Consistent Hashing has significant differences in shard volumes.
\end{itemize}

\section{Discussion}\label{sec:discussion}

Let us now discuss the compliance of each of the considered algorithms with the criteria defined in Section~\ref{sec:evaluation}.
\begin{itemize}
    \item \textbf{Linear Hashing} has proper record distribution among shards and high lookup speed, but it moves up to 80\% of records on each rebalancing step, so this method does not meet our criteria. However, Linear Hashing can be applied in systems with a constant number of shards.
    \item \textbf{Consistent Hashing} has acceptable lookup times while moving an optimal number of records, but distributes data extremely unevenly. Methods with more uniform distribution should be preferred. To improve the distribution quality, one can increase the number of points for each shard on the ring, but this will slow down lookups.
    \item \textbf{Rendezvous} is optimal in case of data rebalancing and distribution, but has the longest lookup time. Since lookup speed is the least prioritized criterion, this method is suitable for us.     
    \item \textbf{RUSH\textsubscript{R}} does not satisfy all three criteria, so it does not suit our goals.
    \item \textbf{Maglev} provides fast lookup and relatively uniform distribution, but in some cases it can move more than 50\% of all records (see the horizontal line on Figure~\ref{fig:moved-etalons}). Therefore, Maglev is appropriate for systems with a fixed number of shards.
    \item \textbf{Jump} moved the largest number of records (Table~\ref{tbl:experiment1}), so it is unsuitable as well.
    \item \textbf{AnchorHash} appears to be the winner so far as it satisfies all the requirements.
\end{itemize}

Following the results of all three experiments, we have created a table where we listed all evaluated algorithms (Table~\ref{tbl:criteria}). We have assessed them according to our three criteria and assigned a rating out of three quality grades~--- low, medium, and high.

\begin{table}[]
\centering
\caption{Load balancer criteria satisfaction table}
\label{tbl:criteria}
\begin{tabular}{l|c|c|c}
Algorithm & \begin{tabular}[c]{@{}c@{}}Data distribution\\ uniformity\end{tabular} & \multicolumn{1}{l|}{Rebalancing quality} & \multicolumn{1}{l}{Lookup speed} \\
\hline \hline
Linear     & High   & Low    & High   \\ \hline
Consistent & Low    & High   & Medium \\ \hline
Rendezvous & High   & High   & Low    \\ \hline
RUSH       & Low    & Low    & Low    \\ \hline
Maglev     & High   & Medium & High   \\ \hline
Jump       & Medium & Low    & Medium \\ \hline
AnchorHash & High   & High   & Medium
\end{tabular}
\end{table}

It is evident from the table that there are two winning algorithms~--- Maglev and AnchorHash, which fail to reach either top rebalancing quality (number of moved records) or the top lookup speed.

AnchorHash distributes data uniformly, moves an optimal amount of records and its lookup time is small enough. Rendezvous also fits first and second criteria, but its lookup time is more than two times larger than that of AnchorHash. These two methods are appropriate for systems with frequent shard addition or removal.

On the other hand, Maglev's lookup is more than two times faster, therefore it is suitable for static systems, similarly to Jump and Linear Hashing.

Consistent Hashing seems to be effective for both types of systems, but its main drawback is non-uniform data distribution among shards.

RUSH\textsubscript{R} has been proven to be the worst algorithm out of all.

\section{Acknowledgments}

We would like to thank Anna Smirnova for her help with the preparation of the paper.

\section{Conclusion and Future Work}\label{sec:concl}

In this paper we have studied several hashing algorithms and assessed their applicability to data balancing in distributed databases. For this, we have performed both simulated and real experiments. The real experiments were run using the Unidata platform, an open-source tookit for building MDM solutions. In these experiments we have employed three criteria of applicability, namely uniformity of produced data distribution, the amount of moved records, and computation costs.

Experiments demonstrated that out of seven considered algorithms there are two clear winners~--- AnchorHash and Maglev. Another two, Linear Hashing and Jump, may have some applicability too.

There are several possible directions for extending this paper. First of all, we noticed some impact of the random number generator on the behavior of some algorithms. In the current paper, we fixed it for all algorithms, but it may be worthwhile to explore this influence. Secondly, it may be interesting to study the impact of varying the parameters of the algorithms. In this paper we have used either the default or the recommended ones, but it is possible that careful tuning may yield positive results. 

\bibliographystyle{splncs04}
\bibliography{my}

\end{document}